\documentclass{aa}
\usepackage{graphicx}
\begin{document}

\title{The transverse velocity and excitation structure of the HH~110 jet 
 \thanks{Based on observations made with the 4.2 m William Herschel Telescope 
operated on La Palma by the Issac Newton Group of Telescopes at the Observatorio del Roque 
de los Muchachos of the Instituto de Astrof\' \i sica de Canarias. } 
}

\author{A.  Riera \inst{1,2} 
\thanks{on sabatical leave at the  Instituto de Ciencias Nucleares, 
Universidad Nacional Aut\'onoma de M\'exico, Apartado Postal 70-543, 04510 M\'exico D.F., M\'exico}
     \and R. L\'opez \inst{2}
     \and A. C. Raga  \inst{3} 
     \and R. Estalella \inst{2}
     \and G. Anglada \inst{4}
      }

   \institute{Departament de F\'\i sica i Enginyeria Nuclear, Universitat
Polit\`ecnica de Catalunya, Av. V\' {\i}ctor Balaguer s/n, E-08800
Vilanova i la Geltr\'u, Spain \\
              \email{angels.riera@upc.es}
         \and
       Departament d'Astronomia i Meteorologia, Universitat de
Barcelona, Av.\ Diagonal 647, E-08028 Barcelona, Spain \\
            \email{robert,rosario@am.ub.es}
        \and 
         Instituto de Ciencias Nucleares, Universidad Nacional Aut\'onoma de M\'exico,
Apartado Postal 70-543, 04510 M\'exico D.F., M\'exico\\
             \email{raga@astroscu.unam.mx}
       \and 
Instituto de Astrof\'\i sica de Andaluc\'\i a, CSIC, Camino Bajo de Hu\'etor
                     24, E-18008 Granada, Spain   \\
              \email{guillem@iaa.es}
}

 \offprints{A. Riera}
\date{Received October 23, 2002; accepted December 16, 2002}
\authorrunning{Riera et al. }
\titlerunning{The transverse structure of the HH~110 jet }

\abstract{We present long-slit spectroscopic observations of the HH~110 jet obtained 
with the 4.2~m William Herschel Telescope. We have obtained for the first time, 
 spectra for slit positions 
along and across the jet axis (at the position of knots B, C, I, J and P) 
to search for the observational signatures of entrainment and turbulence 
by studying the kinematics and the excitation structure. We find that the HH~110 
flow accelerates from a velocity of 35~km~s$^{-1}$ in knot A up to 110~km~s$^{-1}$ 
in knot P. We find some systematic trends for the variation of the emission line ratios 
along the jet. No clear trends for the variation of the radial velocity  
are seen across the width of the jet beam. The cross sections of the jet show complex radial 
velocity and line emission structures which differ quite strongly from each other.
\keywords{ISM: individual (HH~110) --- ISM: jets and outflows ---
stars: pre-main-sequence}
}

\maketitle

% The different journals have different requirements for keywords.  The
% keywords.apj file, found on aas.org in the pubs/aastex-misc directory, 
% contains a list of keywords used with the ApJ and Letters.  These are 
% usually assigned by the editor, but authors may include them in their 
% manuscripts if they wish. 

\section{Introduction}

The HH~110 jet is found in L1647, in the Orion B cloud complex. There is some evidence that
this object corresponds to the early stages of a jet-cloud collision. The jet
presents a rather
complex morphology in H$\alpha$ and [S~II] images (Reipurth \& Olberg 1991), 
with noticeable wiggles
along its length, being suggestive of a turbulent flow. This object shows a very
different morphology in near-infrared images (Davis et al. 1994, Noriega-Crespo et al. 1996).
 In the molecular hydrogen lines, the jet is nearly straight and the molecular emission appears
shifted westward relative to the optical emission. The morphological differences and the
displacement between the optical and near infrared emissions have been interpreted 
as the result of a
grazing collision between the jet and a dense molecular cloud core. In this scenario, the jet
strikes the molecular core and is deflected. The H$_2$ emission is tracing the region where
the atomic and molecular gas interact (Raga \& Cant\'o 1995). This proposed 
scenario is actually
reinforced by observational evidences, and has recently been modeled in some detail
(Raga et al. 2002).

The morphology of HH~110 first suggested
that the driving jet source would be embedded in the dark lane, north of the apex of the flow.
However, searches at optical, near infrared and radio continuum wavelengths
have failed to detect a
driving source along the HH~110 flow axis. Later on, Reipurth et al. (1996)
 reported the discovery of
another jet, HH~270, located $3'$ northeast of HH~110. An embedded near infrared source,
very close to IRAS 05489+0256, is found along the HH~270 flow axis. Based on the morphology
and on the direction of the proper motions of HH~270 (which point toward the apex of HH~110), 
Reipurth et al. (1996)
proposed that HH~270 and HH~110 are associated. The HH~270 flow, driven by the
embedded source, could be having a grazing collision with a
dense molecular clump, through which the jet is deflected into the HH~110 flow
(Raga et al. 2002).

Further observational results also are in agreement with this scenario.
Rodr\'\i guez et al. (1998) detected centimeter 
radio continuum emission at the position of the embedded near infrared source. This radio
continuum source (VLA~1) appears sligthly  elongated in the NE-SW direction, with the major
axis aligned with the HH~270 flow axis. This suggests that the VLA~1 source traces the
base of the flow giving rise to the HH~270/HH~110 system. On the other hand, 
Sep\'ulveda (2001)  
detected, through ammonia mapping, a high density clump of molecular gas which is
spatially associated
with the HH~110 flow, being located in the region where HH~270 abruptly changes
its direction to
emerge as HH~110. This high density gas, traced by the ammonia emission, could be
responsible for the deflection of the HH~270 flow.

Choi (2001) has obtained HCO$^+$ interferometric maps, and does not detect
emission in the crossing point of the HH~270 and HH~110 flow axes. Choi points out
that this result is in apparent contradiction with the ``deflected jet'' interpretation
of these flows. However, Raga et al. (2002) suggest that this result is not necessarily
in contradiction with the deflected jet model, since the impact point might
have moved into the dense cloud, so that it would now lie to the West of the
HH~110 axis (where Choi 2001 does detect extended HCO$^+$ emission).

The HH~110 jet seems to be suitable for searching for the observational signatures 
of entrainment and turbulence (as predicted by theoretical models)
by studying the kinematics and the excitation
structure along and across the jet axis.
In this paper, we present 
the results obtained from long-slit,
high resolution spectroscopy of the HH~110 jet.
We have obtained spectra for slit positions both along and
across the jet beam, allowing us to measure the spatially resolved radial
velocities and emission line ratios  along and across the jet.
The observations are described in Sect.  2, and the results are
presented in Sect.  3. Finally, the results are discussed in Sect.  4.

The data discussed in this paper represent a significant step forward
with respect to previous spectroscopic observations of HH~110. As far
as we are aware, the only published spectra of this object are the
low resolution spectrum of Reipurth \& Olberg (1991) and the high resolution
spectrum of Reipurth et al. (1996), both corresponding to a single slit position
along the outflow axis. We are now presenting high resolution spectroscopy with
substantially more extensive spatial coverage, consisting of several slit positions
both along and across the outflow axis. Throughout the paper, we attempt to
compare our results (at the appropriate spatial positions)
with the ones obtained from the spectrum of Reipurth et al. (1996), and
the line intensities obtained from our spectra with the narrow band
images of Reipurth et al. (1996) and Noriega-Crespo et al. (1996).

\section{Observations}

Long-slit, high resolution spectra of the HH~110 jet were
obtained during the nights of December 12 and 13, 1998. 
The red arm of the double-armed spectrograph ISIS (Carter, Benn, \& 
Rutten  et al. 1994) and a Tektronics CCD detector of $1024 \times 1024$ pixels (pixel size
of 24 $\mu$m) were used at the Cassegrain focus of the 4.2 m William
Herschel Telescope (WHT) at the Observatorio del Roque de los
Muchachos (La Palma, Spain). The high resolution grating R1200R
(dispersion 17 \AA~mm$^{-1}$) 
centered at 6600 \AA~ was employed, covering the H$\alpha$, [N~II] 6548, 
6583 \AA~ and [S~II] 6717, 6731 \AA~  lines and providing a spectral
 sampling of 0.41 \AA~ pixel$^{-1}$ (equivalent to 20 km s$^{-1}$ 
at H$\alpha$). The spatial sampling was 0.36$''$~pixel$^{-1}$ and the seeing
had a FWHM of $2''$ to $3''$. The spectrograph slit had a projected width of $1''.5$.

\begin{figure*}
   \centering
   \includegraphics[width=8cm]{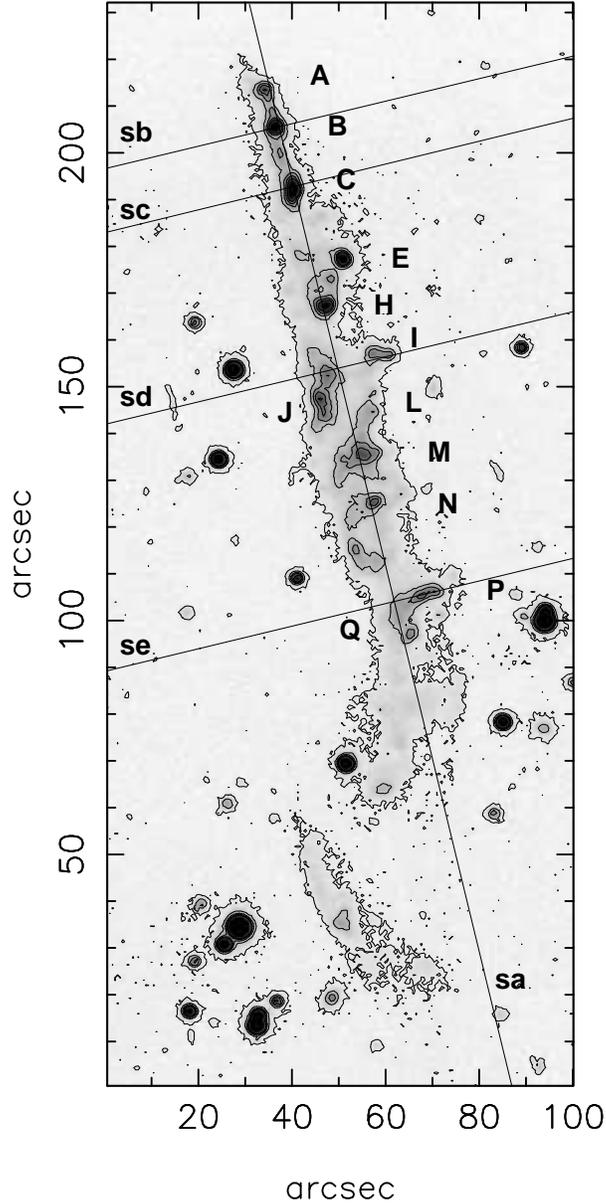}
   \caption{[S~II] 6717+6731 image of HH~110 obtained by the authors on Dec. 1993 at the 2.5~m
   Isaac Newton Telescope (INT) at the Observatorio del Roque de los Muchachos. 
   On this image, the slit positions used for the
   spectroscopy described in the paper have been indicated. North is up and East is to
   the left.}
              \label{image}
    \end{figure*}

Spectra of the HH~110 jet were obtained  at different positions: 
in one of these positions, the slit was placed along the central
 axis of the jet. 
For the rest of the spectra, the slit was positioned perpendicular
 to the jet axis and each of these spectra was obtained cutting the 
jet beam across a bright knot (see Table 1 and
Fig.~\ref{image}). This means that, for the first time, we were able to study 
the kinematic and excitation structure across the flow. 

\begin{table}
\centering
\caption[ ]{Journal of Observations}
\begin{tabular}{lcccc}
\hline
Date & Knot & P.A. & Slit Position$^1$ & Exp. Time \\
     &      & (degrees) &   &  (sec) \\
\hline
%\startdata  
Dec 12 & A-Q &  13.6 & sa & 1800 \\
       & B   & 103.6 & sb & 3600 \\
       & C   & 103.6 & sc & 3600 \\
       & I-J & 103.6 & sd & 3600 \\
Dec 13 & A-Q &  13.6 & sa & 1800 \\
       & I-J & 103.6 & sd & 1800 \\
       & P-Q & 103.6 & se & 5400 \\
\hline
$^1$ See Fig.~\ref{image}
\end{tabular}
\label{tlogin}
\end{table}

The spectra were reduced using the standard tasks for long-slit 
spectroscopy within the IRAF package\footnote{IRAF is distributed by the National Optical 
Astronomy Observatories, which is operated by the Association of Universities for 
Research in Astronomy, Inc., under a contract agreement with the National Science 
Foundation}, that includes bias substraction,
 flat-fielding correction, wavelength calibration and
sky substraction. The spectra were corrected for cosmic ray events
 by median filtering several
exposures obtained at the same slit position. The spectra were
 not flux calibrated.

The results obtained from these observations are described in the
following section.

\section{Results}

\subsection{Structure along the HH~110 jet}

\begin{figure*}
   \centering
   \includegraphics[width=17cm]{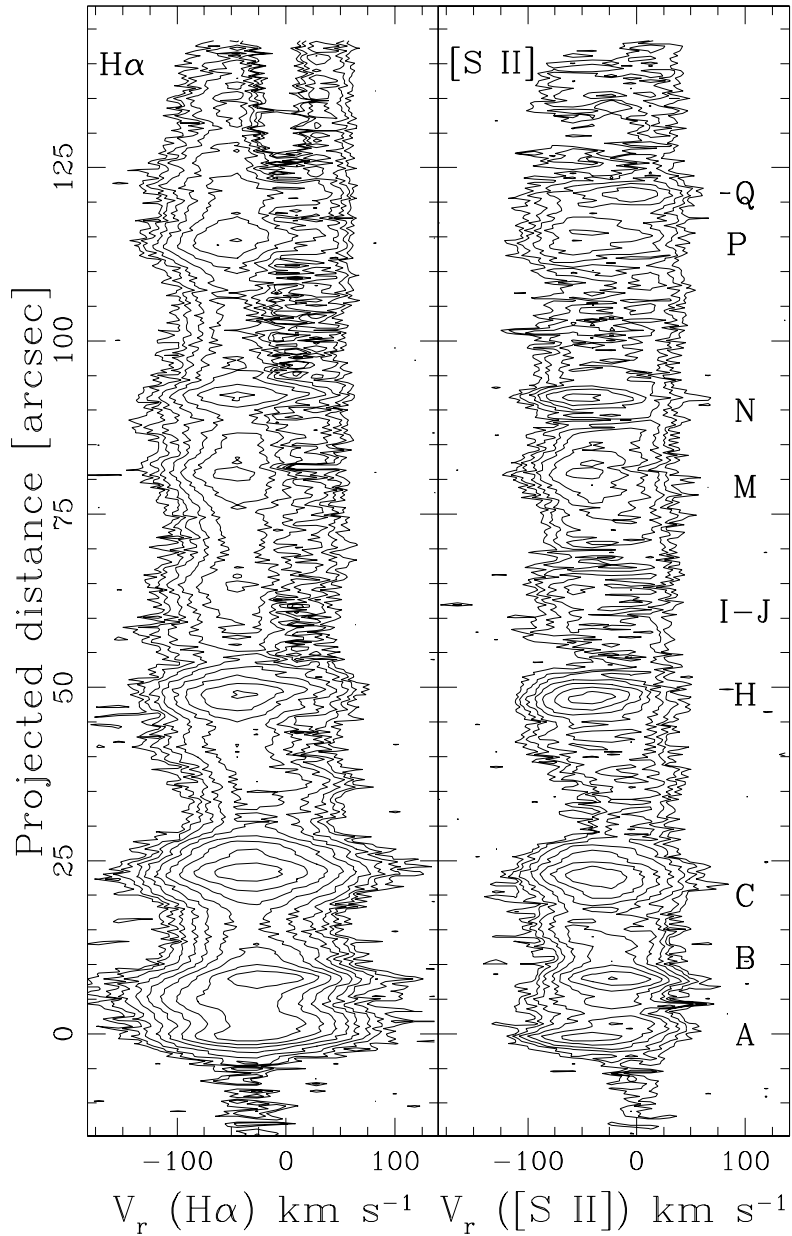}
   \caption{H$\alpha$ and [S~II] 6717 \AA~ long slit spectra
along the axis of the HH~110 jet (slit position ''sa'' in Fig. 1), 
depicted with factor of 2$^{1/2}$, logarithmic contours. 
The radial velocities are relative to the ambient molecular cloud,  
which has a heliocentric radial velocity of 23~km~$^{-1}$  
(see Reipurth \& Olberg 1991). The distances are measured from knot A}
              \label{velal1}
    \end{figure*}

In Fig.~\ref{velal1}, we present the H$\alpha$ and [S~II] 6717 \AA~
PV (position-velocity) diagrams along the jet axis.
We have used the nomenclature of Reipurth \& Olberg (1991) for the knots,
the most prominent of which are identified in Fig.~\ref{velal1}.
The radial velocities are relative to the ambient molecular cloud 
which has a heliocentric radial velocity of 23~km~s$^{-1}$ (Reipurth \& Olberg 1991). 
Our PV diagrams are consistent with the [S~II] PV diagram of
Reipurth et al. (1996).

  \begin{figure}
    \resizebox{\hsize}{!}{\includegraphics{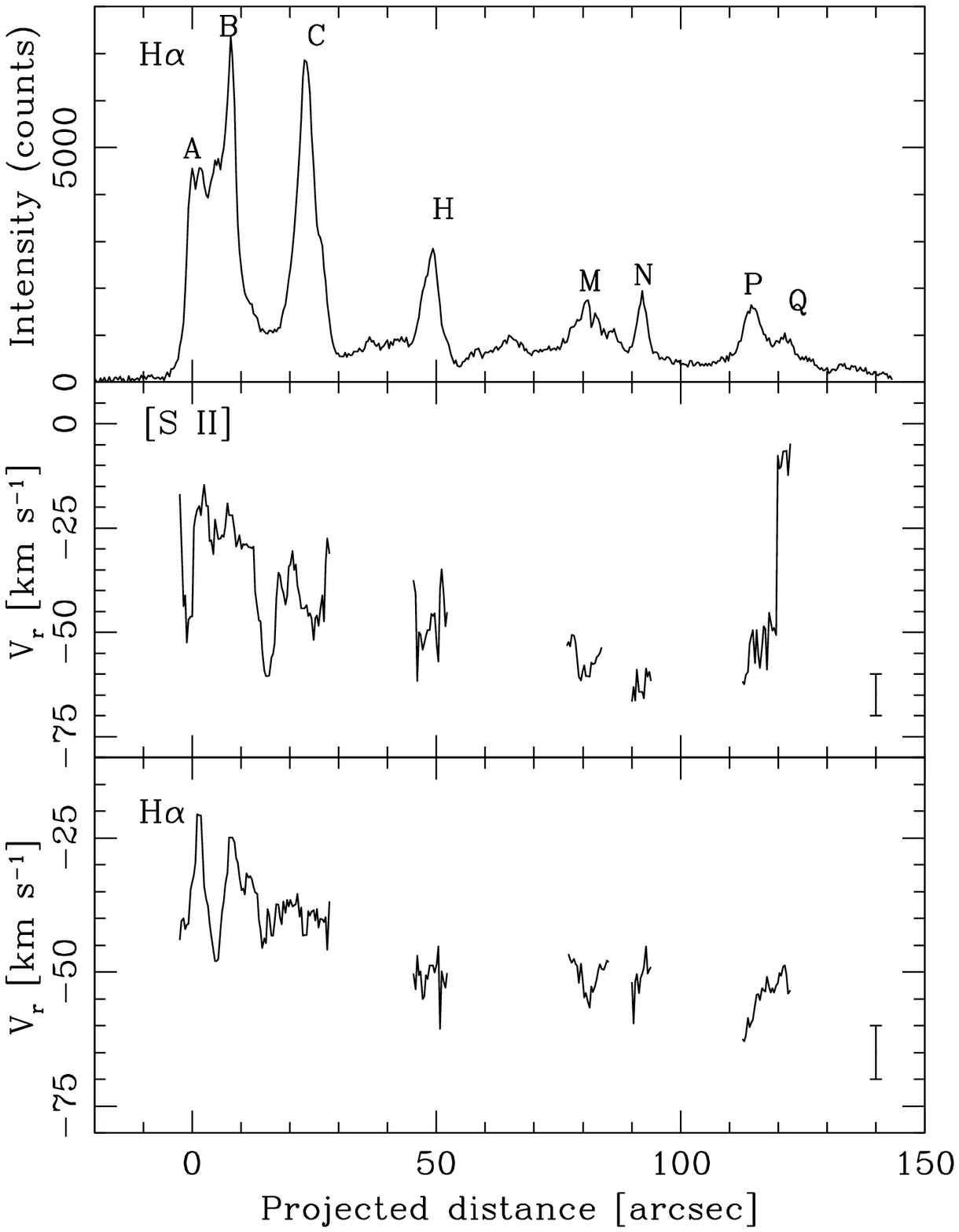}}
    \caption{Radial velocity corresponding to the peak of the H$\alpha$ ({\it bottom})
and [S~II] 6717 \AA~ line profiles
({\it center}) as a function of position along HH~110. Only the regions with
a signal-to-noise ratio which is high enough to obtain a reliable determination
of the radial velocities of the peak emission are shown.
The radial velocities shown are relative to the ambient molecular cloud. 
The error bars in the central and bottom panels indicate typical errors in the
radial velocities.
The H$\alpha$ spatial intensity distribution is also shown ({\it top})
in order to help with the identification of the knots. The distances are measured
from knot A.}
              \label{velal2}
    \end{figure}

Fig.~\ref{velal1} shows a change in radial velocity 
along knot A, with a decrease of 25~km~s$^{-1}$ (from $-$48 to $-$23~km~s$^{-1}$) 
in $1\farcs7$. Such an effect was already noted by Reipurth et 
al. (1996). 
From Fig.~\ref{velal1}, we also see that the FWHM of the [S~II] 6717 \AA~ 
emission line has the highest values at knots B and C, 
with a value of $\sim$ 90 km s$^{-1}$, and decreases to a value of 70 km s$^{-1}$ 
at knot H, in clear agreement with the results of Reipurth et al. (1996). 
Beyond knot H, the line width remains approximately constant. 
We should note that these line widths are well resolved with our 
30~km s$^{-1}$ effective spectral resolution.

Fig.~\ref{velal2} shows the radial velocities (with respect to the molecular cloud)
corresponding to the peaks of
the H$\alpha$ and [S~II] 6717 \AA~ line profiles along the HH~110 jet.
All knots show blueshifted emission lines, with relatively small radial velocity
variations along the axis of the jet.

The radial velocity of knot B is of $\sim$ $-$25~km~s$^{-1}$, and it increases (in modulus)
at the position of knot C to $\sim$ $-$43~km~s$^{-1}$ in [S~II] (see Fig.~\ref{velal2}). 
Knot H has a radial velocity of $\sim$ $-$50 km s$^{-1}$.
At the location of knot M, the [S~II] and H$\alpha$ lines show a range of
radial velocities, between $\sim$ $-$50~km~s$^{-1}$ and $-$60~km~s$^{-1}$. 
Knot N has a radial velocity of $\sim$ $-$55~km~s$^{-1}$. 
Knot P shows a range of radial velocities from $-$65 to $-$55~km~s$^{-1}$.

Finally, knot Q has a radial
velocity of $-$50~km~s$^{-1}$ in H$\alpha$ and $-$10~km~s$^{-1}$ in [S~II].
This difference appears to be a result of the fact that at the position of knot
Q, two components with different radial velocities are seen in both the
H$\alpha$ and [S~II] line profiles. While in H$\alpha$ the high radial velocity
component is dominant, in the [S~II] lines the low velocity component is much
brighter, resulting in the low radial velocity determined from the [S~II] lines.

Fig.~\ref{velal2} reveals that the variation of radial velocities with distance 
along HH~110 is qualitatively similar (with the exception of knot Q)
for the H$\alpha$ and [S~II] emission lines. In both cases, the jet accelerates 
with distance along the jet from $-$20 to $-$65~km~s$^{-1}$ (with respect to
the molecular cloud).

\begin{figure}  
   \resizebox{\hsize}{!}{\includegraphics{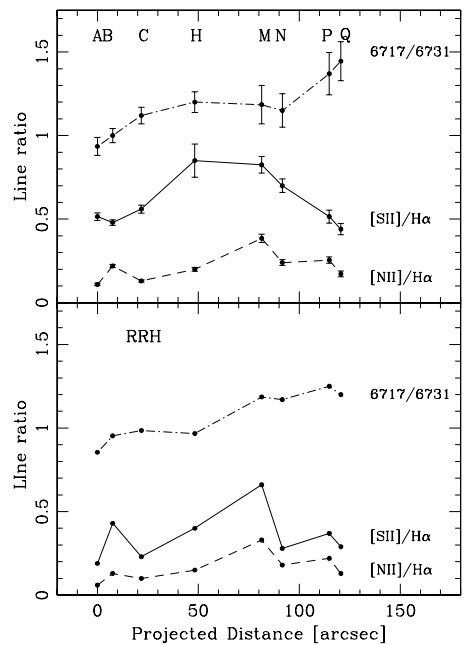}}
   \caption{The [N~II] 6583~/~H$\alpha$,  
the [S~II]~(6717+6731)~/~H$\alpha$, and the [S~II] 6717~/~6731 ratios 
computed from the velocity integrated line intensities
are plotted for the knots intercepted by the slit position ``sa''.  
The top panel corresponds to the present data 
(where the two spectra obtained along the axis of the jet were averaged),  
and the bottom panel has been taken from Reipurth et al. 
(1996, labeled as RRH). }
              \label{ratal} 

    \end{figure}

\begin{figure}
   \resizebox{\hsize}{!}{\includegraphics{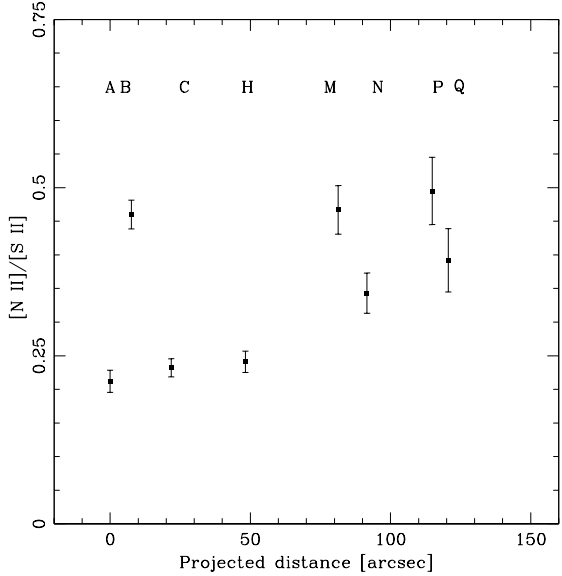}}
   \caption{The velocity integrated  [N~II]~6583~/~[S~II]~(6717+6731)
line ratio is plotted for the knots intercepted by the slit position ``sa''. 
 }
              \label{niisii} 

    \end{figure}

In Fig.~\ref{ratal}, we present the velocity integrated line ratios for various knots,
as derived from the long-slit spectra. The  [N~II] 6583~/~H$\alpha$, the [S~II] 
(6717+6731)~/~H$\alpha$, and the [S~II] 6717~/~6731 emission line ratios 
are plotted as a function of position along the slit. 
 The emission line ratios obtained at December 12 and 13 are similar to each
other (implying that the slit positions probably were almost coincident).
In Figures ~\ref{ratal} and  ~\ref{niisii} we show the average
of both spectra (with error estimates based on the statistical error
as well as on the differences between the two observations),
and the following discussion is also based on the average of both spectra.
For comparison, we have also included the results from Reipurth et al.
(1996). 

The [N~II] 6583~/~H$\alpha$ emission line ratio has a value of $\sim$ 0.12
in the bright knot A. This ratio increases  in knot B, and decreases further out at the 
position of knot C. This ratio is $\sim$ 0.20 at the position of knot H, and it 
increases to $\sim$ 0.40 at knot M. At the position of knots N and  P  
the [N~II]~/~H$\alpha$ ratio has an approximately constant value
 ($\sim$ 0.25) and decreases beyond knot P. 

The trend of variation of the [N~II] 6583~/~H$\alpha$ emission line ratio with angular distance 
is consistent with the results of Reipurth et al. (1996). However, the values reported 
by Reipurth et al. (1996) are significantly lower than our values.

The overall trend of variation of the [S~II]~/~H$\alpha$ 
emission along the length of the slit shows an almost constant value for the 
brightest knots (knots A, B, and C), where the ratio
is $\sim$ 0.50. The [S~II]~/~H$\alpha$ emission line ratio rapidly increases
along knots H and M, to values of $\sim$ 0.80. The [S~II]~/~H$\alpha$ 
ratio decreases at larger distances (at the position of knots N, P and Q). 

Fig.~\ref{ratal} illustrates 
the differences between our results and the [S~II]~/~H$\alpha$ 
ratio trend along the HH~110 jet derived by Reipurth et al. (1996). 
We observe a rapid increase of the [S~II]~/~H$\alpha$ ratio from knot B to the
location of knots H-M, and a linear decrease beyond 
knot M, while Reipurth et al. (1996) obtained a slow increase from knot B
to knot M and a rapid decrease beyond knot M. Moreover, our [S~II]~/~H$\alpha$ 
ratios are significantly higher than the values reported by Reipurth et al.
(1996).

Fig.~\ref{ratal} also shows the [S~II] 6717~/~6731 emission line ratios along 
the HH~110 jet. A first glance at the [S~II] 6717~/~6731  plot shows a rapid 
increase  of this ratio (and, consequently, a rapid decrease of the  
electron density) with increasing distance along the jet. The electron density 
decreases from $\sim$ 865 cm$^{-3}$ in knot A to $\sim$ 640  cm$^{-3}$ in knot B,
with a lower value of $\sim$ 415 cm$^{-3}$ in knot C. The [S~II] 6717~/~6731 ratio
increases at knot H, and consequently the electron density decreases to a 
value of $\sim$ 270 cm$^{-3}$. The electron density increases at knots M and N 
(with values of $\sim$ 500 cm$^{-3}$). The other knots show smaller electron 
densities ($\sim$ 150 cm$^{-3}$ in knot P and $\sim$ 50 cm$^{-3}$ in knot Q). 

The overall behaviour of the electron density along the jet axis is
qualitatively similar to the measurements of Reipurth et al. (1996), who
also found an almost continous decrease of the electron density along the
axis of HH~110. However, our electron density values 
are somewhat lower than the values reported by Reipurth et al. (1996). 

Finally, Fig.~\ref{niisii} illustrates the behaviour of the [N~II]~6583~/~[S~II]~(6717+6731)
(hereafter [N~II]~/~[S~II]) ratio (which is a measure of the excitation of the
spectrum) along the axis of HH~110. The [N~II]~/~[S~II] ratio shows 
a monotonic increase with distance from a value $\sim$ 0.20 (at knot A) to $\sim$ 0.50 
(at knots P and Q), with the exception of knot B which shows a
[N~II]~/~[S~II] line ratio of $\sim$ 0.45, which lies well above the prevailing
trend.

\subsection{Structure across the HH~110 jet: knots B and C}

First, we study the PV diagrams and the variation of the
[N~II] 6583~/~H$\alpha$ and [S~II] (6717+6731)~/~H$\alpha$
emission line ratios and the
electron density as a function of position across the jet at knots B and C, which lie
closer to the base of the HH~110 jet than the other knots, and where the jet shows a
well collimated morphology (see Fig.~\ref{image}). 

\begin{figure}
   \resizebox{\hsize}{!}{\includegraphics{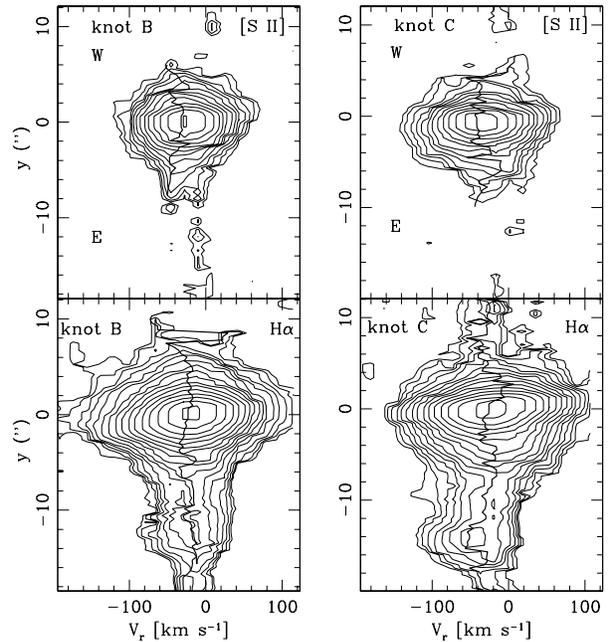}}
   \caption{H$\alpha$ ({\it bottom panels}) and [S~II] 6717 \AA~
({\it top panels}) long slit spectrum across the jet at the positions of
knots B ({\it left panels}) and C ({\it right panels}); depicted 
with 2$^{1/2}$, logarithmic contours. The superimposed solid line is the
radial velocity (with respect to the molecular cloud)
corresponding to the peak of the line profile. 
}
              \label{velbc} 

    \end{figure}

In Fig.~\ref{velbc}  we present the PV diagrams across 
the beam of the jet. These measurements were obtained with spectrograph slit
positions ``sb''  and ``sc'' (see Fig.~\ref{image} and Table 1), which intersect 
the jet at knots B and C, respectively. The distances were measured from the 
H$\alpha$ intensity peak (located at 0.0 in the Fig.~\ref{velbc}).  
We should remember that the molecular cloud is located towards the west 
(i.e., positive values of $y$). 

The H$\alpha$ and [S~II] emission lines of knots B and C 
are characterised by the presence of a peak and a weak asymmetrical 
emission which extends eastwards. In the PV
diagrams of knots B and C (see Fig.~\ref{velbc}), 
we clearly see that the [S~II] emission extends out to 6$''$ from the peak of the knot
towards the west, and to $\sim$ 9$''$ towards the east.  
We also see that the H$\alpha$ emission extends 10$''$ further out to the east.
The secondary emission we have detected in H$\alpha$ corresponds to the faint
emission detected east of knots B and C in the H$\alpha$
image obtained by Reipurth \& Olberg (1991, who labeled this region as X and Y).

The H$\alpha$ and [S~II] 6717 \AA~ PV diagrams of knots B and C show
an approximately constant radial velocity across the knots (see Fig.~\ref{velbc}). 
The radial velocities (i.e. the mean values of the bright central regions of 
knots B and C)  have values of  $\sim$ $-$30~km~s$^{-1}$. 
These radial velocities  are consistent within the errors with the velocity 
values derived for knots B and C from the spectra obtained along the axis 
of the jet (see Sect.  3.1). We see an increase of the radial velocity (in modulus) by  
20~km~s$^{-1}$ from the central part of knot B towards the edges 
(see Fig.~\ref{velbc}), while no significant variation is observed across knot C.

\begin{figure}
   \resizebox{\hsize}{!}{\includegraphics{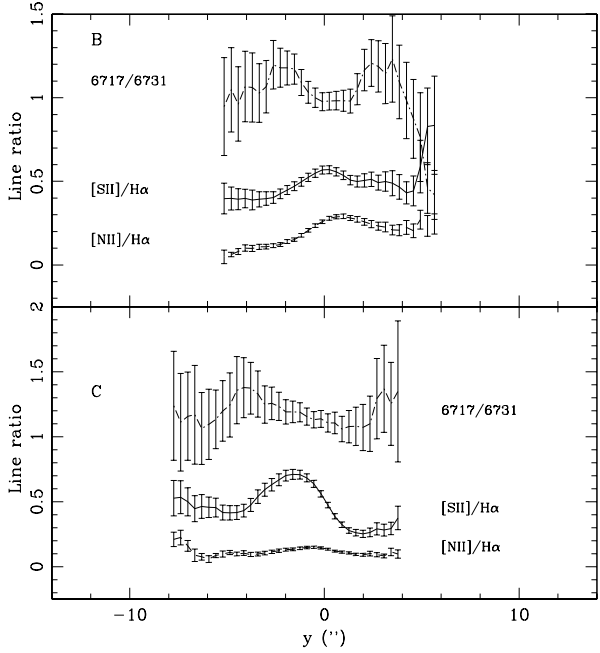}}
   \caption{[N~II] 6583~/~H$\alpha$,   [S~II] (6717~+~6731)~/~H$\alpha$, and [S~II] 
6717~/~6731 line ratios  through the cross section
 of the jet at the positions of knots B ({\it top panel}), and C
 ({\it bottom panel}), as a function of distance from the central peak.
 The intensities were integrated over all the radial velocity range in
 which line emission was detected. }
              \label{ratiosbc} 

    \end{figure}

Across knot B from west  to east, the [N~II] 6583~/~H$\alpha$ emission line ratio first
increases from 0.20 to 0.30 in $\sim$ 3$''$ 
and then rapidly decreases (see Fig.~\ref{ratiosbc}). The maximum of the
[N~II]~/~H$\alpha$ ratio is displaced by $\sim$ $0\farcs8$ towards the west
with respect the central peak of the H$\alpha$ intensity.  

The [S~II] (6717~+~6731)~/~H$\alpha$ ratio shows a 
similar, but less pronounced, variation across knot B (see Fig.~\ref{ratiosbc}). 
This ratio first increases 
from $\sim$ 0.40 to 0.55 in $\sim$ 4$''$, and then decreases eastwards to a
value of $\sim$ 0.40 at $-$4$''$ from the central position. 

Knot B has an almost constant [S~II] 6717~/~6731 
line ratio of $\sim$ 0.95 (see Fig.~\ref{ratiosbc}) across the central bright region, 
corresponding to a $n_e$ $\sim$ 700 cm$^{-3}$. The 
density decreases at $y\sim \pm$ 2$''$, where [S~II] 6717~/~6731 $\sim$ 1.17 (e.g.,  
$n_e$ $\sim$ 300 cm$^{-3}$). 

The [N~II]~/~[S~II] line ratio decreases across knot B from $\sim$ 0.60 at 2$''$ west of the 
central peak (i.e. $y=0$) to  $\sim$ 0.25 at 4$''$ east of the central peak.

Across knot C, the [N~II] 6583~/~H$\alpha$ emission line ratio is significantly
lower than the corresponding values for knot B, which is in 
agreement with the results obtained from the spectrum along the axis of the jet.
The [N~II]~/~H$\alpha$ ratio slightly increases 
moving eastwards across the bright, central region (see Fig.~\ref{ratiosbc}), 
and then decreases further estwards. A larger variation across knot C was found
for the [S~II] (6717+6731)~/~H$\alpha$ ratio, which increases 
across the central region (see Fig.~\ref{ratiosbc}) and then decreases further
eastwards. The maximum of the [S~II]~/~H$\alpha$ 
distribution is located at $\sim$ $1\farcs4$ east of the H$\alpha$ central peak.  

At the central position of knot C, the [S~II] 6717~/~6731 ratio is $\sim$ 1.08
(see Fig.~\ref{ratiosbc}), and therefore the electron density has a value of 
$\sim$ 450 cm$^{-3}$. The [S~II] 6717~/~6731 ratio increases towards both sides,
adopting a value $\sim$ 1.35 at $\sim \pm$ 4$''$ from the centre (i.~e., 
$n_e$ $\sim$ 100 cm$^{-3}$). The [N~II]~/~[S~II] emission line ratio decreases across
knot C from west to east, from $\sim$ 0.40 to 0.20 across 5$''$.

The results found in this section therefore are:
\begin{itemize}
\item the radial velocities of knots B and C are approximately constant across
the width of the jet,
\item the [N~II]~/~H$\alpha$ and [S~II]~/~H$\alpha$ ratios show an increase across 
knots B and C moving from west to east (along $\sim$ 6$''$),
\item the electron densities are constant in the brightest region of knots B and C,
which are surrounded by lower density gas,
\item the [N~II]~/~[S~II] emission line ratios across knots B and C show a similar
trend, with a west to east decrease (i.e., a decrease of the excitation of the spectrum).   
\end{itemize}

\subsection{Structure across HH~110: knots I, J, P and Q}

At a  position of 25$''$ south of  knot A, the jet widens in a cone with
a large opening angle. The slit positions ``sd'' and ``se'' (see Fig.~\ref{image}
and Table 1) show emission extending $\sim 30''$ in 
[S II] cutting across knots I and J (slit position ``sd''), 
and also across knot P (slit position ``se''). The spectrum across knot P
also shows emission arising from the northern edge of knot Q (see Fig.~\ref{image}).

\begin{figure}
   \resizebox{\hsize}{!}{\includegraphics{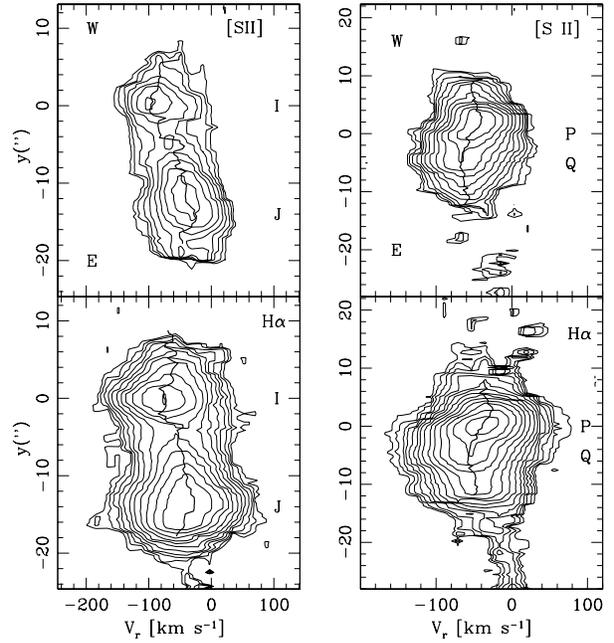}}
   \caption{H$\alpha$ ({\it bottom panels}) and [S~II] 6717 \AA~ 
({\it top panels}) long slit spectrum 
across the jet at the position of knots I+J ({\it left panels}) and P+Q
({\it rigth panels}); depicted with 2$^{1/2}$, logarithmic contours.
The superimposed solid lines are the radial velocities (with respect to
the molecular cloud) corresponding to the peak of the line profiles. 
The angular distances are measured from the central H$\alpha$ peak of knot I
({\it left panels}) and knot P ({\it right panels}).
}
              \label{velijpq} 

    \end{figure}

In Fig.~\ref{velijpq} we present the PV diagrams  
across the beam of the jet at the position of knots I and J, and P and Q. 
Across knot J, we see a clear trend of decreasing
radial velocities (in modulus) moving 
from west to east. In the western edge of knot J we measure a
[S~II] 6717 \AA~ radial velocity of $-$55~km~s$^{-1}$, and at a position $\sim$ 9$''$ to the 
east this velocity has a value of $\sim$ $-$25~km~s$^{-1}$.
In knot I, the peak of the [S~II] 6717 \AA~ emission has a velocity
of $\sim$ $-$70~km~s$^{-1}$, and the H$\alpha$ emission has a radial velocity
of $\sim$ $-$85~km~s$^{-1}$.

Fig.~\ref{velijpq} also shows the PV diagram across knots P and Q.
The H$\alpha$ radial velocity as a function of position shows an S-shaped structure,
with a peak value (in modulus) of $-$67~km~s$^{-1}$ and a minimum value
of $-$40~km~s$^{-1}$. The peak of knot P (located at $y=0$) has a radial velocity
of $-$55~km~s$^{-1}$. The [S~II] radial velocities have a similar dependence with
position, but have values which are systematically less negative by 10~km~s$^{-1}$.

\begin{figure}
   \resizebox{\hsize}{!}{\includegraphics{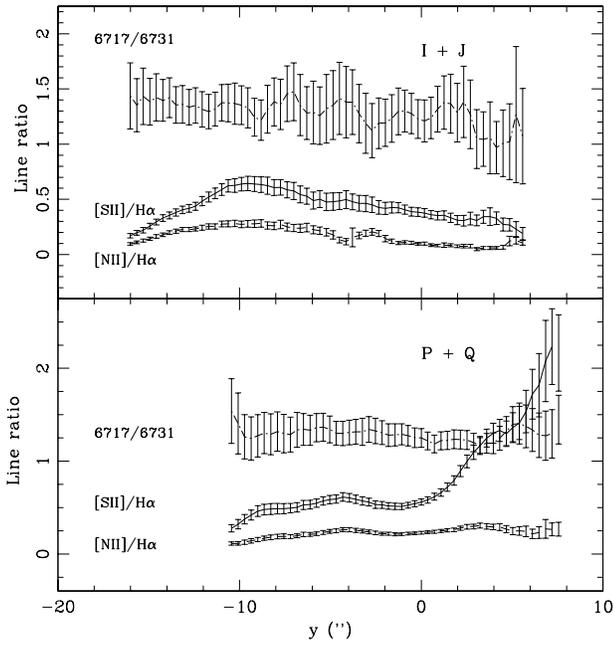}}
   \caption{[N~II] 6583~/~H$\alpha$, [S~II] (6717+6731)~/~H$\alpha$, and [S~II] 
6717~/~6731 line ratios  through the cross section
 of the jet at the positions of knots J + I ({\it top panel}), and P + Q
 ({\it bottom panel}), as a function of distance from the central peak.
 The intensities were integrated over all the radial velocity range in
 which line emission was detected. }
              \label{ratiosijpq} 

    \end{figure}

Fig.~\ref{ratiosijpq} shows the [N~II] 6583~/~H$\alpha$, [S~II] (6717+6731)~/~H$\alpha$, and 
[S~II] 6717~/~6731 emission line ratios across knots J+I, and P+Q. 
The [N~II] 6583~/~H$\alpha$ ratio increases as a function of distance across 
knots I and J (from west to east).  The [N~II] 6583~/~H$\alpha$ ratio increases 
across knot I (from west to east) up to a value of 
$\sim$ 0.30. The [N~II] 6583~/~H$\alpha$ ratio rapidly decreases (from west to east)
across knot J (see Fig.~\ref{ratiosijpq}), decreasing from 0.30 to a value of 0.075 
along $7\farcs0$. 

The [S~II] (6717+6731)~/~H$\alpha$ line ratio also shows a linear increase across 
knot I and across the gas between the two knots (increasing from
 0.30 to 0.65 over 12$''$, see Fig.~\ref{ratiosijpq}). 
From the western to the eastern edge of knot J,
the [S~II]~/~H$\alpha$ ratio decreases from 0.6 to 0.2.   

The [N~II]~/~[S~II] emission line ratio changes abruptly at the position of the H$\alpha$ 
peak of knot I, showing a value of $\sim$ 0.20 to the west and $\sim$ 0.40 to
the east. At the position of knot J, the [N~II]~/~[S~II] ratio has a constant
value of $\sim$ 0.45.

The [S~II] 6717~/~6731 line ratio shows an overall increasing trend
from west to east (see Fig.~\ref{ratiosijpq}). The electron density has a mean 
value $\leq$ 100 cm$^{-3}$ at the position of knot I. 
We see a W-E trend of increasing [S II] 6717/6731 line ratio across knot J, 
with electron densities decreasing from $\sim$ 100  cm$^{-3}$ to $<$ 60 cm$^{-3}$.

Fig.~\ref{ratiosijpq} also illustrates the variation of the [N~II] 
6583~/~H$\alpha$ ratio as a function of position across knot P. 
Along the western edge of knot P, from west to east, the [N~II]~/~H$\alpha$ ratio
shows two peaks. The higher peak is displaced
$\sim 3''$ to the west of the central part of knot P. The lower peak approximately
coincides with knot Q (even though this knot does not lie right on the position
of our spectrograph slit, see Fig.~1).

The [S~II]~/~H$\alpha$ ratio rapidly decreases from west to east across knot P
from a value $>$ 1.3 (at the western edge) to 0.7
(at the location of the H$\alpha$ maximum, see Fig.~\ref{ratiosijpq}). 
Across knot Q, the [S~II]~/~H$\alpha$ ratio is $\sim$ 0.9, 
and smoothly decreases to the east of the knot.

The [S II] 6717~/~6731 ratio increases across knot P (from west to east), 
with values of $\sim$ 1.25 to 1.35 along 6$''$ (see Fig.~\ref{ratiosijpq}) .  
The electron densities range from  200 cm$^{-3}$ to 80 cm$^{-3}$.  
The [N~II]~/~[S~II] emission line ratio first increases eastwards,
and then maintains a more or less constant value of $\sim 0.40$ across the
brighter regions of knots P and Q.

The results found in this section therefore are:
\begin{itemize}
\item the radial velocities of knots I and J decrease from west to east, while 
the PV diagram of knot P shows a S-shaped structure, 
\item the [N~II]~/~H$\alpha$ and [S~II]~/~H$\alpha$ ratios increase across 
knots I and J moving from west to east, and show an overall decreasing trend across 
knot P, 
\item the electron densities decrease from from west to east across knots I, J and P,
\item the [N~II]~/~[S~II] emission line ratios abruptly change at the position of knot I 
(from a value of $\sim$ 0.20 west to 0.40 east of the knot), and this ratio is more or 
less constant across the bright regions of knot P.
\end{itemize}

\subsection{ The spatial displacements between the H$\alpha$ and [S~II] spatial
intensity distributions}

Noriega-Crespo et al. (1996) pointed out the differences observed between 
the intensity distribution of the optical emission (H$\alpha$ and [S~II]) and
the H$_2$ emission along and across the axis of the HH~110 jet. 
An inspection of their Figs. 3, 4 and 5 also shows clear differences between 
the H$\alpha$ and the [S~II] emission lines. 

\begin{figure}
   \resizebox{\hsize}{!}{\includegraphics{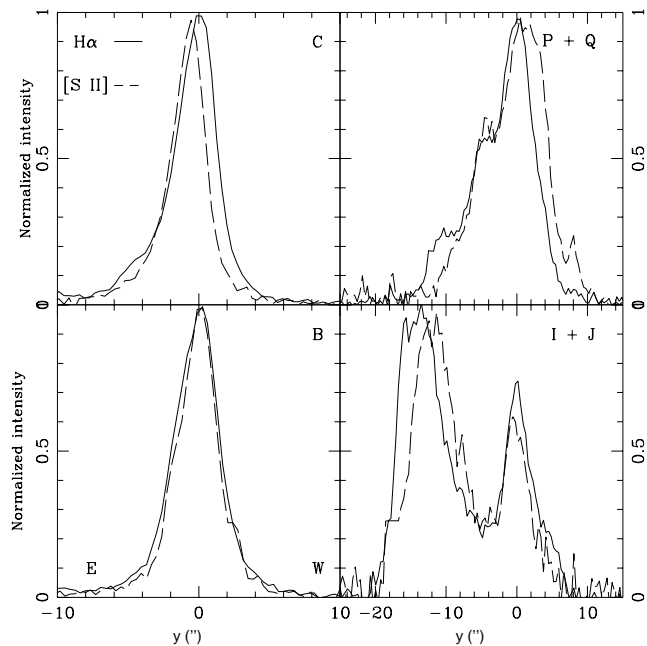}}
   \caption{Spatial H$\alpha$ ({\it solid line}) and [S~II] ({\it dashed line}) 
intensity distributions through the cross section of the jet at 
the location of knots B, C, I+J and P+Q. The zero point of the spatial axis 
is defined in each knot at the position of the peak intensity 
of H$\alpha$.  The distributions have been normalized 
by their peak intensities. 
}
              \label{perf} 
\end{figure}

In the following, we explore the differences between the spatial distributions
of the H$\alpha$ and [S~II] velocity integrated intensities (see 
Fig.~\ref{perf}) across the axis of the HH~110 jet, as obtained from the spectra
with slit positions ``sb''-''se'' (see Table 1). 

In  Fig.~\ref{perf} (top left panel) we show the H$\alpha$ and [S~II] intensity 
distribution over the cross section of the jet at knot C. 
This graph reveals the apparent spatial separation between both distributions. 
We observe a clear offset in position of the intensity peaks, with the 
[S~II] intensity maximum at $0\farcs5$ towards the east with respect to the
H$\alpha$ peak. For knot B, we observe a similar spatial distribution
in H$\alpha$ and [S~II], in agreement with the results obtained by
Noriega-Crespo et al. (1996). 

For knot P, we find a displacement between the H$\alpha$
and the [S~II] 6717 \AA~ spatial intensity distributions
(see Fig.~\ref{perf}, top right panel). We observe a clear offset 
(by $0\farcs8$) between the position of the intensity maximum 
seen in H$\alpha$ and in
[S~II]~6717 \AA. In contrast to the spatial displacement observed 
across knot C, the [S~II] peak is displaced towards
the west with respect to the H$\alpha$ peak, again in agreement with
the results of Noriega-Crespo et al. (1996). 

Finally, across knot J we observed the largest spatial offset between the 
H$\alpha$ and [S~II] spatial intensity distribution  (see Fig.~\ref{perf}, 
bottom right panel), while no offset is observed across knot I. 
Across knot J, the [S~II] peak is offset by $2\farcs0$
westwards from the H$\alpha$ peak.

The values of the observed displacements between the H$\alpha$ and 
the [S~II]~6717 \AA~ spatial intensity distributions should be taken with some 
caution due to the $\sim 2-3''$
angular resolution of our observations. However, we do find a good agreement 
between the spatial displacements shown in our spectra  and 
the results of Noriega-Crespo et al. (1996). This agreement appears to
confirm the reality of the spatial displacements which we see in our spectra.

\section{Discussion and Conclusions}

As we have described in the previous sections, the HH~110 jet has a very
complex structure, with detailed, spatially dependent radial velocities,
line intensities and line ratios. Let us now describe the general trends
that can be deduced from the detailed structure.

The result that is most clear is that the radial velocity has a
largely monotonic trend of larger (i.~e., more negative) velocities
as a function of position along the jet (see Sect.  3.1). The velocity goes from
a value of $-20$~km~s$^{-1}$ for knot A to $-65$~km~s$^{-1}$ in knot P.
If we consider that the jet has a mean radial velocity (with respect
to the cloud) $<v_r>=-42$~km~s$^{-1}$, one can use the mean proper motion
$<v_p>=83$~km~s$^{-1}$ determined by Reipurth et al. (1996), we obtained
a $\phi=27^\circ$ between the outflow axis and the plane of the sky.
On the other hand, if we consider the radial velocities and proper motions of the
H, M and P knots (which are the knots with the better determined
proper motions) to obtain an inclination angle $\phi_H=22^\circ$,
$\phi_M=41^\circ$ and $\phi_P=38^\circ$, respectively. Interestingly, averaging these
three values we obtain an angle of $34\pm 8^\circ$ which is consistent within
the errors with the $27^\circ$ orientation angle calculated with the mean
radial velocity and proper motion (see above).

Adopting a $\phi=35^\circ$ angle between the outflow axis and the plane
of the sky, we can deproject the measured radial velocities, and conclude
that the HH~110 flow accelerates from a velocity of 35~km~s$^{-1}$ in
knot A up to 110~km~s$^{-1}$ in knot P (knot A and P being separated
by $130''$). This acceleration could be the result of a time-dependent
ejection velocity from the outflow source (with ejection velocity decreasing
as a function of time). Another possible explanation for such a velocity vs.
position trend has been suggested by Raga et al. (2002), who modeled HH~110
as a wake left behind by a short-lived deflected jet, which is pinched
off as the incident jet starts to penetrate the dense cloud which
produced the jet deflection.

The radial velocity cross sections of the jet (see Sects. 3.2
and 3.3), show an almost
constant velocity for knots B and C, a velocity change from W to E
of $\sim 60$~km~s$^{-1}$ (corresponding to a deprojected value
of $\sim 100$~km~s$^{-1}$) for knots I and J, and a more complicated,
S-shaped cross section spanning a $\sim 30$~km~s$^{-1}$
velocity range (corresponding to a deprojected value of
$\sim 50$~km~s$^{-1}$) for knot P.

The [S~II] line ratio along the HH~110 jet shows a more or less monotonic
drop of the electron density in the downstream direction (see Sect.  3.1).
The [S~II]~/~H$\alpha$ ratio first grows, reaching a maximum of
$\sim 0.8$ in knots H-M, and then drops along the jet axis. The
[N~II]~/~H$\alpha$ ratio has a qualitatively similar behaviour,
showing a maximum in knot M. The [N~II]~/~[S~II] ratio shows
a more or less monotonic growth down the jet axis.

The [S~II] line ratio
across the jet (see Sect.  3.3) shows that across knots B and C
we have a dense ($\sim 450-700$~cm$^{-3}$), central region of the
jet beam, with lower density ($\sim 100-300$~cm$^{-3}$)
regions on both sides. In knots I, J and P, the electron density
grows monotonically from E to W (from $\sim 50$ to 200~cm$^{-3}$).
The other measured line ratios show quite complex cross sections.

If we compare the cross sections of the emission of the different
lines, we find that the maxima sometimes coincide (knot B and I),
but that in other cases there are shifts of $~\sim 0\farcs5$-$2''$ between
the maxima of H$\alpha$ and [S~II] (knots C, J and P, see Sect.  3.4).

From the above discussion, we see that while there are systematic
trends along the HH~110 jet for both the radial velocities and the emission
line ratios, no clear trends are seen across the width of the
jet beam. The cross sections of the jet (taken
at different positions along the HH~110 jet) show complex radial
velocity and line emission structures, which differ quite
strongly from each other.

Given the complex structures that are observed, it does not appear
to be reasonable to compare them with simple models of plane-parallel
shock waves, or with ``mixing-length'' models of turbulent mixing 
layers. For example, the analytic turbulent mixing layer model of Noriega-Crespo
et al. (1996) predicts that there should be a systematic velocity increase
from W to E across the section of the HH~110 flow. Also, the model predicts
that there should be a systematic W to E increase in the excitation of the
emitted spectrum. Such systematic trends might be partially present
in our data, but they mostly show complex cross sections which differ from each other
at different positions along the HH~110 flow. For example, we do see that
the [N~II]~/~[S~II] ratio increases from W to E, and we detect spatial offsets
between the maxima of the H$\alpha$ and [S~II] emission cross sections,
in qualitative agreement with the mixing layer model of Noriega-Crespo et al.
(1996). However, we do not see clear trends in the radial velocity across
the HH~110 flow.

We should note that one might expect to have
large discrepancies between observations of a turbulent flow (which would
correspond to a ``snapshot'' of the turbulent eddies) and the predictions
from a ``mixing length'' turbulent mixing layer model (which attempts
to describe the properties of a smooth, ensemble-averaged ``mean flow'').
Therefore, it is difficult to use comparisons between observations and
this type of models to prove
or disprove whether the emission from HH~110 is indeed formed in
a turbulent mixing layer.

A possible approach would be to compute 3D models of a jet/cloud
interaction similar to the ones of Raga et al. (2002), but including
a description of the S~II and N~II ions, so as to be able to obtain
predictions of the observed emission lines. From such a simulation,
one could explore whether or not the complex properties which
we have discussed above are consistent with a jet/cloud collision
scenario.

As noted by Raga et al. (2002), a model of a jet/cloud collision does
produce a ``pinched-off'' deflected jet beam that has properties which resemble the
H$\alpha$ and H$_2$ 1-0 s(1) emission of HH~110. Such a model also
produces an apparent ``acceleration'' along the flow, which is the
characteristic signature of the wake left behind the pinched-off
deflected jet beam. Our data do show this acceleration, which is also seen
in the proper motions of Reipurth et al. (1996).

The spectroscopic
data presented in this paper, however, give a much more detailed description
of the kinematics and excitation of HH~110. It will be necessary to compute new
models (with higher resolution and a more detailed treatment of the
atomic/ionic processes) in order to see whether or not these observations agree
in detail with the predictions from a jet/cloud interaction model.

\begin{acknowledgements}
The work of RE, RL and AR was supported by CGICYT grant PB 98-0670-C02 
and MCyT grant AYA2002-00205 (Spain). The work of GA was   
supported by CGICYT grant PB 98-0670-C02 (Spain). 
RE and AR  acknowledges support from the ACI 2000-8 
Direcci\'o General de Recerca (Spain).
GA acknowledges support from Junta de Andaluc\'\i a (Spain). 
The work of ACR was supported by CONACyT grants 34566-E and 36572-E.  
\end{acknowledgements}


\begin{thebibliography}{}



\bibitem[1994] {car94} Carter, D., Benn, C.R., Rutten, R.G.M.,  Breare, J.M., Rudd,  P.J., 
King, D.L., Clegg, R.E.S., Dhillon, V.S., Arribas, S., Rasilla, J.-L., Garc\'\i a, A.,
  Jenkins, C.R. \& Charles, P.A. 1994, WHT ISIS user's Manual,
 Isaac Newton Group Telescopes, Royal Greenwich Observatory.

\bibitem[2001]{Choi}Choi, M. 2001, ApJ, 550, 817  

\bibitem[1994] {dav94} Davis, C.J., Mundt, R. \& Eisl\"offel, J. 1994, ApJ, 437, L55. 

\bibitem[1996] {nor96} Noriega-Crespo, A., Garnavich, P.M., Raga, A.C., Cant\'o, J., 
\& B\"ohm, K.-H., 1996, ApJ, 462, 804.
 
\bibitem[1995] {rag95}  Raga, A.C. \& Cant\'o, J. 1995, Rev. Mexicana Astron. Astrof. 31, 51.

\bibitem[2002]{rag02} Raga, A.C., de Gouveia Dal Pino, E.M., Noriega-Crespo, A., 
Mininni, P.D. \& Vel\'azquez, P.F. 2002, A\&A, 392, 267

\bibitem[1991] {rei91} Reipurth, B.\& Olberg, M., 1991, A\&A, 246, 535.

\bibitem[1996] {rei96} Reipurth, B., Raga, A.C. \& Heathcote,
 S., 1996, A\&A, 311, 989.
 
\bibitem[1998] {rod98} Rodr\' \i guez, L.F., Reipurth, B., Raga, A.C. \& Cant\'o J. 1998,
Rev. Mexicana Astron. Astrof., 34, 69.  
 
\bibitem[2001] {sep01} Sep\'ulveda, I., 2001, Ph.D. Thesis, U. Barcelona 

\end{thebibliography}
\end{document}